\documentclass[letter]{ptptex}
\usepackage{wrapft}

\usepackage{graphicx}

\pubinfo{Vol.~120, No.~5, November 2008} 

\title{
Effects of a new triple-$\alpha$ reaction rate on the helium ignition of  accreting white dwarfs
}

\author{
Motoaki \textsc{Saruwatari}$^{1,}$\footnote{
E-mail: saruwatari@phys.kyushu-u.ac.jp} 
and Masa-aki \textsc{Hashimoto}$^{1,}$\footnote{
E-mail: hashimoto@phys.kyushu-u.ac.jp}
}

\inst{
$^1$Department of Physics, Kyushu University, Fukuoka 812-8581, Japan
}

\abst{%
Effects of a new triple-$\alpha$ reaction rate on the  ignition of carbon-oxygen white dwarfs accreting helium  in a binary systems have been investigated.  The ignition points determine the properties of a thermonuclear explosion of a Type Ia supernova. We examine the cases of different accretion rates of helium and different initial masses of the white dwarf,
   which was studied in detail by Nomoto\cite{rf:nom82}. We find that  for all cases from slow to intermediate accretion rates, nuclear burnings are ignited at the helium layers. As a consequence, carbon deflagration would be triggered for
 the lower accretion rate compared to that of $dM/dt\simeq 4\times10^{-8} M_{\odot}~\rm yr^{-1}$ which
 has been believed to the lower limit of the accretion rate for the deflagration supernova.
   Furthermore, off-center helium detonation should result for intermediate and slow accretion rates and the region of carbon deflagration for slow accretion rate is disappeared.  
}

\PTPindex{242, 421}   

\begin{document}
\maketitle

\noindent
1. {\it Introduction\qquad}
Triple-$\alpha$ (3$\alpha$) reaction plays an important role for the helium burning stage on the stellar evolution of low, intermediate and high mass stars\cite{rf:sn80,rf:nh88,rf:hashi95}, and accreting white dwarfs\cite{rf:nom82,rf:ntm95}. Recently, the 3$\alpha$ reaction has been calculated by Ogata et al.\cite{rf:okk}, which is very large compared with the previous rate used so far.\cite{rf:ntm95,rf:angulo,rf:dotter} It should be examined how the new rate affects the astrophysical phenomena, because terrestrial experiments for the 3$\alpha$ reaction are very difficult such as the study of the QCD phase transition at high densities.   We  investigate the effects of  a newly calculated 3$\alpha$ reaction rate (OKK rate)\cite{rf:okk} on the helium flashes, which are occurred at the center or
inside layers in the accreting envelope of the compact stars.
The ignition curves  play a critical role when  the nuclear burning  begins to occur and  becomes the main energy source to change the stellar structure, where the fates of the  massive stars and/or accreting white dwarfs are determined by the strength of specified nuclear burnings.\cite{rf:nom82,rf:hashi86} While nuclear burning depends on the temperature severely, the density becomes very important at high density of
$\rho \geq 10^6~\rm g~cm^{-3}$ and low temperature of $T \leq 10^8$~K, because the screening effects begin to enhance the reaction rates. 

  It has already shown that the new rate affects significantly the   evolutionary tracks of low mass stars  for 1 and 1.5  $M_{\odot}$, where the red giant phase  has  almost disappeared.  As the result, they concluded that the rate is not  compatible with observations.\cite{rf:dotter}   We can see the effects of the OKK rate on the  stellar evolution with use of the ignition properties. The helium core  flash is triggered if the nuclear generation rates ($\varepsilon_n$) overcome the neutrino loss rates ($\varepsilon_{\nu}$) significantly.   Figure 1 shows the  ignition curves of  two 3$\alpha$ reactions and the evolutionary tracks of the central temperature ($T_c$) against the central density ($\rho_c$)
  for stars from 1.5 to 40 $M_{\odot}$ in the main sequence stages. The evolutions are
  calculated beyond the core helium burning with the previous 3$\alpha$ reaction rate.
 We can understand clearly that the helium ignition occurs in the considerably  low temperature and density points compared to the previous case.\cite{rf:sn80,rf:tny84}

     \begin{figure}
         \centerline{\includegraphics[width=10 cm,height=15 cm,angle=-90]
                                     {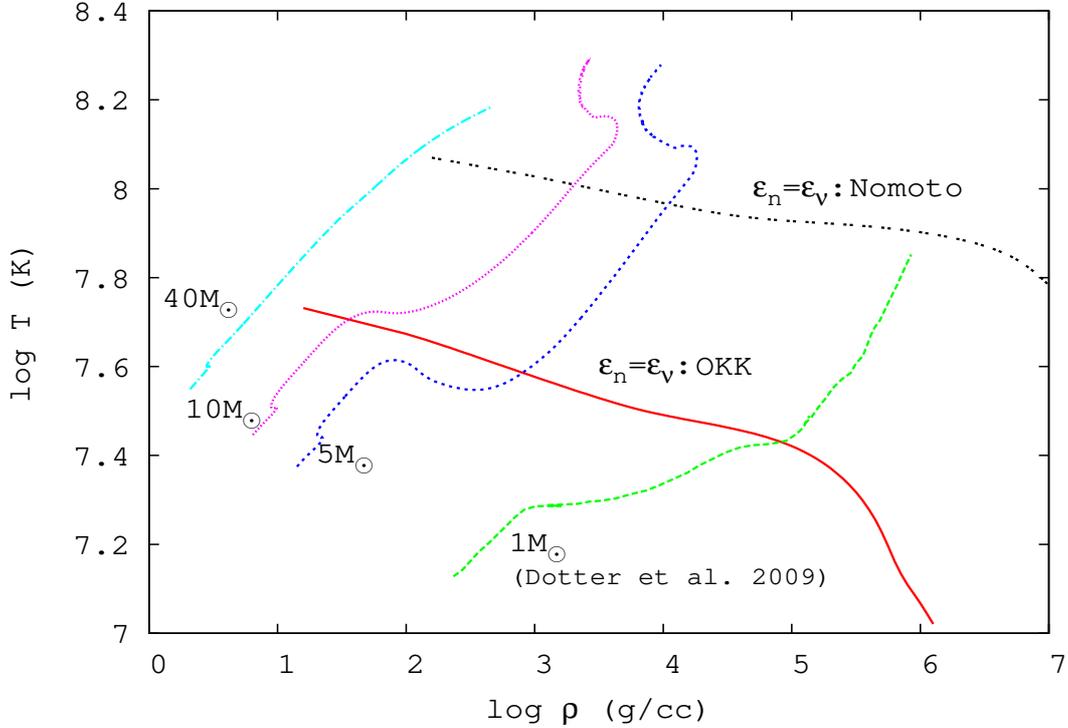}}
     \caption{Two ignition curves of $\varepsilon_n =\varepsilon_{\nu}$ are
     obtained from the previous rate (dashed line) and OKK rate (solid line),
     respectively.
     Evolutionary tracks of ($\rho_c, T_c$) with the previous 3$\alpha$ reaction
rate are shown for the stars of 1-40 $M_{\odot}$. For the star of 1$M_{\odot}$, the track is taken from Ref.~\citen{rf:dotter}. For other stars, the evolutions are calculated from the zero-age main sequence stage.\cite{rf:saio,rf:yamaoka}}
     \label{fig:1}
     \end{figure}

\noindent
2. {\it Ignition curves and helium flash on the accreting white dwarfs\qquad}
Accreting white dwarfs are considered to be  the origin of the Type Ia supernova explosions.\cite{rf:tny84}  While the  white dwarfs are composed mainly of carbon and oxygen (CO),  accreting materials are  usually hydrogen and/or helium.  Since the hydrogen is converted to  helium through steady hydrogen burning, helium is accumulated on  the white dwarf gradually and the deep layers become hot and  dense.   Helium  flash  triggered in the regions composed of degenerate  electrons  could develop to  the dynamical stage, which depends  on the accretion rate $dM/dt$.\cite{rf:tny84} The properties of ignition depend on the initial mass of the white dwarf
$M_{\rm C+O}$ for slow accretion rates. If the accretion is rather rapid, $dM/dt\geq 4\times10^{-8} M_{\odot}~\rm yr^{-1}$, the
 carbon deflagration supernova has been considered to be triggered at the center.\cite{rf:nom84} 
  Figure 2 shows the ignition curves concerning the helium flash with the old and new 3$\alpha$ reaction rates adopted, which gives ($\rho_c, T_c$) for
  the beginning of the helium flashes.  Nomoto\cite{rf:nom82} found that the ignition occurs  on the condition of $\tau_n = 10^6$~yr, where the time
  scale of the temperature increase by the nuclear reaction
   is defined to be $\tau_n = C_p~T/\varepsilon_n$  ($C_p$ is the specific heat  at constant pressure).  Evolutionary tracks of ($\rho_c, T_c$) for   cases A--F are taken from
      the figures in Nomoto\cite{rf:nom82} which can be used until the helium   flash begins: (cases A--F: $M_{\rm C+O}~(M_{\odot}), dM/dt~(M_{\odot}~\rm yr^{-1})$); (case A: 1.08, $3\times10^{-8}$), (case B: 1.08, $3\times10^{-9}$), (case C: 1.28 , $7\times10^{-10}$)
(case D: 1.35 , $7\times10^{-10}$), (case E: 1.13, $4\times10^{-10}$),
(case F: 1.28, $4\times10^{-10}$).
We can find that the helium ignitions occur  in the low density  by  almost two  orders  of magnitude if the OKK rate is adopted.    Contrary to the results  in Ref.~\citen{rf:nom82}, nuclear flashes are triggered for all cases of A--F  in the helium  layers  which  are accumulated on the CO white dwarfs.

     \begin{figure}
         \centerline{\includegraphics[width=10 cm,height=15 cm,angle=-90]
                                     {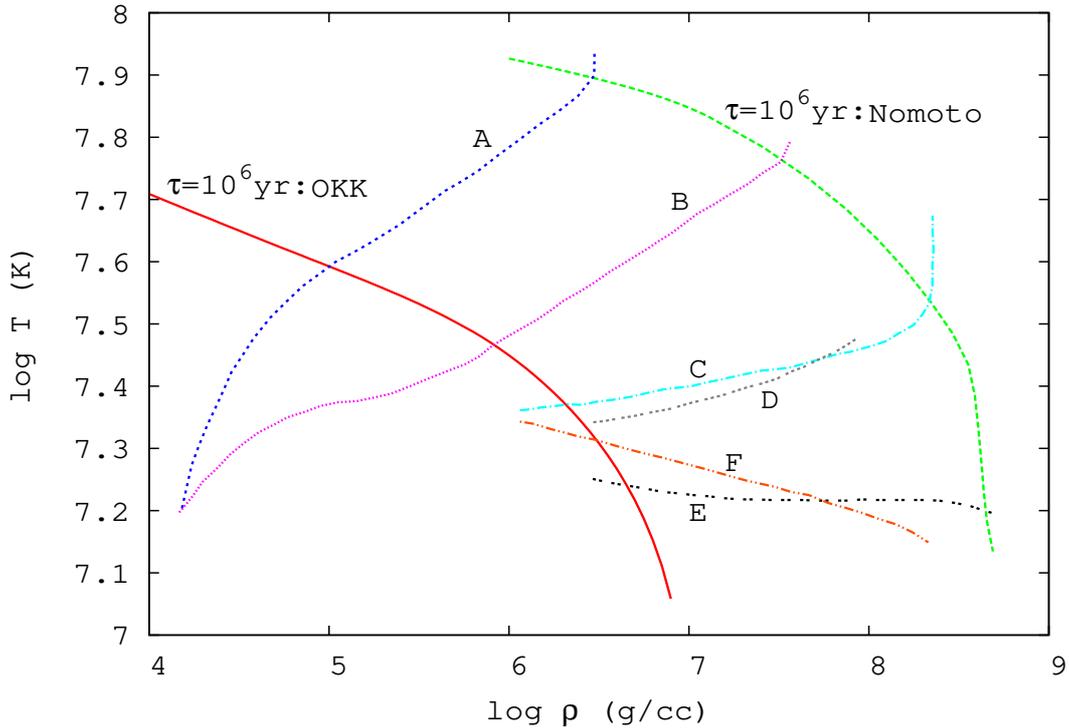}}
     \caption{Ignition  curves defined by $\tau_n = 10^6$~yr for the helium flashes on the accreting white dwarfs with use of the two 3$\alpha$ reaction rates. Evolutionary tracks of ($\rho_c, T_c$) are
     taken from Ref.~\citen{rf:nom82}}
     \label{fig:2}
     \end{figure}

3. {\it Discussion and Conclusions\qquad}
The ignition densities that determine the triggering mechanism
 of Type Ia supernovae  will be changed drastically if we adopt a new  3$\alpha$ reaction rate.  Although Nomoto\cite{rf:nom82} has shown that the specific nonresonant 3$\alpha$ reaction  is crucial in determining the helium ignition density for accretion as slow as 
$dM/dt \leq 10^{-9} M_{\odot}~\rm yr^{-1}$, microscopic calculation for three body problem is found to be much more important to evaluate the 3$\alpha$ reaction rate. Classification due to the accretion rate on the $dM/dt-M_{\rm C+O}$ plane by  Nomoto\cite{rf:nom82} and  Nomoto et al.\cite{rf:ntm95} will be changed significantly.
It was found that when the density in the burning shell become higher than
$2\times10^6~\rm g~cm^{-3}$, the nuclear flash grows into a detonation or deflagration.
We emphasize that the accretion rates which induce the carbon deflagration supernova become much lower compared to the standard rate of $dM/dt\simeq 4\times10^{-8} M_{\odot}~\rm yr^{-1}$.\cite{rf:nom82,rf:ntm95,rf:sn80,rf:tny84,rf:nom84} 
Contrary to the previous calculations, off-center helium detonation dominates  the mechanism of Type Ia supernovae for the low helium accretion rate of $dM/dt \leq 7\times10^{-10} M_{\odot}~\rm yr^{-1}$ without the carbon ignition at the center.  The new rate affects also the helium ignition in accreting neutron stars. In particular, for lower accretion rates, helium burns at lower densities and 
 temperatures, which could change the epoch of a formation of a helium detonation wave and a modeling of Type I X-ray bursts.\cite{rf:ntm95}

\section*{Acknowledgments}
We would like to thank S. Fujimoto for information about the new 3-$\alpha$ reaction and relevant references. We appreciate H. Saio for the permission to use his stellar evolution code.
This work has been supported in part by a Grant-in-Aid for Scientific Research (19104006, 21540272) of the Ministry of Education, Culture, Sports, Science and Technology of Japan.


\begin{thebibliography}{99}


\bibitem{rf:nom82} 
		K. Nomoto, \AJ{253,1982,798}. 
\bibitem{rf:sn80} 
		D. Sugimoto and K. Nomoto, Space Sci. Rev. \ \textbf{25} (1980), 155.
\bibitem{rf:nh88}
		K. Nomoto and M. Hashimoto, Phys. Rep. \textbf{163} (1988), 13.
\bibitem{rf:hashi95}
		M. Hashimoto, Prog. Theor. Phys. \textbf{94} (1995), 663. 
\bibitem{rf:ntm95} 
		K.~Nomoto, F.-K.~Thielemann and S.~Miyaji,
		Astron, Astrophys. \ \textbf{149} (1995), 239.

\bibitem{rf:okk} 
		K. Ogata, M. Kan and M. Kamimura, \PTP{122,2009,1005}.\\
		K. Kan et al. JHP-Supplement-20, 1996, p. 204; 
		K. Kan, Master thesis 1995, in Kyushu University, unpulbished.
\bibitem{rf:angulo} 
		C. Angulo et al. Nucl. Phys. A \ \textbf{656} (1999), 3.
\bibitem{rf:dotter} 
		A. Dotter and B. Paxton, Astron. Astrophys. \ \textbf{507} (2009), 1617.
\bibitem{rf:hashi86}
		M. Hashimoto, K. Nomoto, K. Arai and K. Kaminisi, \AJ{307,1986,687}.
\bibitem{rf:tny84} 
		F.-K. Thielemann, K. Nomoto and Y. Yokoi, Astron. Astrophys. \ \textbf{186} (1984), 644.
\bibitem{rf:saio}
		H. Saio, K. Nomoto and K. Kato, \JL{Nature,334,1988,508}.
\bibitem{rf:yamaoka}
		H. Yamaoka, H. Saio, K. Nomoto and K. Kato, IAU Symposium 143, Wolf Rayet Stars and Intterrelations with Other Massive Stars in Galaxies, 1990, p. 571.  
\bibitem{rf:nom84} 
		K. Nomoto, F.-K. Thielemann and Y. Yokoi, AJ{286,1984,644}.
 
\end{thebibliography}
\end{document}